\documentclass[12pt]{iopart}
\usepackage{graphicx,wrapfig,epsfig,epsf}
\usepackage{amssymb}

\def\pp{p+p\,}
\def\pA{p+A\,}

\def\PbPb{Pb+Pb\,}
\def\pPb{p+Pb\,}

\def\dAu{d+Au\,}

\def\AuAu{Au+Au\,}

\begin{document}

\title[The physics potential of proton-nucleus collisions at the TeV scale]
{The physics potential of proton-nucleus collisions at the TeV scale}

\author{Carlos A. Salgado\footnote{E-mail:carlos.salgado@usc.es}}

\address{Departamento de F\'\i sica de Part\'\i culas and IGFAE\\
Universidade de Santiago de Compostela\\
E-15782 Santiago de Compostela, Galicia-Spain}
\begin{abstract}
The LHC brings nuclear collisions to the TeV scale for the first time and the first data show the qualitative differences of this new regime.The corresponding phase-space available encompasses completely uncharted regions of QCD in which high-density or high-temperature domains can be identified. Proton-nucleus runs are essential for a complete interpretation of the data and for the study of new regimes dominated by large occupation numbers in the hadronic wave function. I comment here the physics opportunities for \pPb runs at the LHC and \dAu runs at RHIC and the corresponding needs in view of the new \PbPb data from the LHC.
\end{abstract}


\section{Introduction}

Proton-nucleus collisions serve a dual purpose in the experimental programs with high-energy nuclear beams \cite{Salgado:2011wc}.  On the one hand, the {\it cold nuclear matter} background is benchmarked by colliding systems where a hot and dense medium is not expected to be produced. Especially relevant for this benchmarking are the studies of nuclear parton distribution functions (nPDFs) --- see \cite{Eskola:2009uj}-\cite{Kovarik:2010uv} for the most recent NLO analyses. On the other hand, new physics domains of QCD, characterized by large gluonic densities in the hadron wave function, are expected to be more easily accessible by increasing the atomic number of the colliding objects. Proton-nucleus collisions provide the best possible tool for such studies before a lepton-ion collider is built.

Proton-nucleus collisions have been experimentally studied at various energies, from CERN SPS ($\sqrt{s}\simeq 17-27$ AGeV) and Fermilab fixed target experiments ($\sqrt{s}\simeq 39$ AGeV) to BNL RHIC ($\sqrt{s}=200$ AGeV). For the last one, deuteron nuclei, instead of protons, were accelerated due to technical reasons. Most of SPS or Fermilab results refer to quarkonia  and Drell-Yan production with nuclear targets. At RHIC several different observables have been studied, in particular in the forward region, looking for clear signals of saturation of partonic densities  --- see below.

Similarly to these previous experiments, the need of proton-nucleus runs at the LHC has been recognized and studies of feasibility \cite{Salgado:2011wc,ref:EPAC2006} and detector performance  \cite{Carminati:2004fp,Alessandro:2006yt,D'Enterria:2007xr,atlas}  carried out. The collision of asymmetric systems at the LHC presents the additional constraints derived from the two-in-one magnet design. From the feasibility of \pA collisions this is especially relevant during injection and ramp while from the purely experimental conditions, it implies rapidity shifts of the center-of-mass reference frame.  Experimental  checks of feasibility will be performed  in fall 2011, having in mind the possibility of a first \pPb run in 2012 before the long shutdown of 2013-2014.

\begin{figure}[h]
\begin{center}
\includegraphics[width=0.5\textwidth]{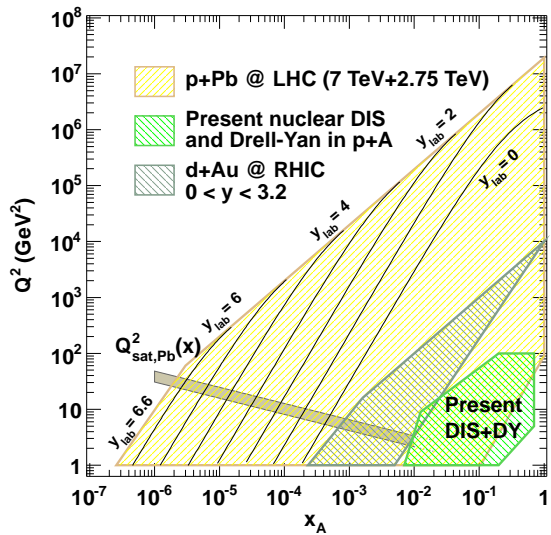}
\end{center}

\caption{Total kinematical reach of \pPb collisions at $\sqrt{s}=$8.8 TeV  at the LHC for different rapidities in the laboratory frame. Also shown are the region of phase space studied by experiments of DIS with nuclei and Drell-Yan production in proton-nucleus collisions (the two main processes used in global nPDF fits) and the total reach of RHIC for $0<y<3$.}
\label{fig:pA}
\end{figure}

The kinematic domains studied by the different experimental facilities is plotted in Fig. \ref{fig:pA}. It should be noticed that the actual reach is normally limited by the kinematics of the process under study and smaller than the one plotted. Interestingly, the central rapidity region at RHIC had already been explored in previous experiments of deep inelastic scattering with nuclei and Drell-Yan production in proton-nucleus collisions. For this reason a check of the universality of the nPDFs is possible. RHIC has also explored the forward region, where smaller values of $x$ are possible. The access, for the first time in nuclear collisions, of the TeV scale, make the reach of the LHC much wider than any previous accelerator. In particular, values of $x$ of the order of $10^{-5}$, well within the saturation region estimates, can be attained with actual processes. Interestingly a large overlap between the LHC and the RHIC kinematics is present. This is very important for cross-checks and for unraveling possible competing cold nuclear matter effects in the data.

It is probably worth mentioning here that the proposed upgrades of the different collaborations \cite{upgrades} emphasize the physics of the forward rapidities, in which a saturation of partonic densities could appear. This physics is best studied in proton-nucleus or deuteron-nucleus collisions.

\section{Some considerations for the LHC as a proton-nucleus collider}

The two-in-one magnet design of the LHC implies an equal rigidity for the accelerated beams. This fixes the momentum of the nucleus to be  \cite{Salgado:2011wc,ref:EPAC2006} 
\begin{equation}
p_{\rm Pb}=Z\,p_{\rm proton}
\end{equation}
with $Z=82$ the number of protons in the Pb nucleus. In general, for the collision of two asymmetric systems $(Z_1,A_1)$ and $(Z_2,A_2)$ one has
\begin{equation}          
\label{eq:rootsy}
\sqrt{s_{\rm NN}}  \approx 2c\,p_{\rm proton} \sqrt {\frac{{Z_1 Z_2 }}{{A_1 A_2 }}}
,\quad \quad
\Delta y  \approx  \frac{1}{2}\log \frac{{Z_1 A_2 }}{{A_1 Z_2 }}
\end{equation}
which leads for the top LHC energy to $\sqrt{s_{\rm pPb}}\simeq 8.8$ ATeV, $\Delta y\simeq$ 0.46. The corresponding energy if the first \pPb run takes place in 2012 is about a factor of two smaller while the rapidity shift remains unchanged.

The corresponding estimates for the luminosity of a \pPb run is
\begin{equation}                    
 \label{eq:typicalLuminosity}
L \approx 1.5 \times 10^{29} {\rm{ cm}}^{{\rm{ - 2}}} {\rm{s}}^{{\rm{ - 1}}} .
\end{equation}
The final luminosity will, of course, depend on a number of factors in which the previous experience on the LHC collisions in both \pp and \PbPb modes will be of special relevance. The figure quoted above is for the top LHC energy (it would be slightly smaller for smaller energy) but takes a rather conservative factor of 10 less intensity than nominal for the proton bunches \cite{Salgado:2011wc,ref:EPAC2006}, so, larger values could be possible. Assuming a total running period of $10^6\,s$ this leads to an integrated luminosity of 0.1 pb$^{-1}$. This luminosity allows to measure hard processes with large virtualities, for example jets are produced copiously, but also rarer processes as $W/Z$ production with enough statistics to be of use in global analyses.

\section{RHIC: Benchmarking and studies of saturation}

The first collisions of deuteron-gold took place at RHIC during Run 3 in 2003. The corresponding results are essential for the interpretation of the \AuAu data, in particular for the interpretation of {\it jet quenching} as a final state effect and also, to a less extent to calibrate the cold nuclear matter effects in $J/\Psi$ production. At the same time, the first studies of particle production at forward rapidities started to point to the interpretation of saturation of partonic densities as a plausible one for the strong suppression observed. 

Concerning the central rapidities, as it was mentioned, the range of $x$ and $Q^2$ measured by high-$p_T$ inclusive pion production overlaps to a large extent with that from previous data on DIS and Drell-Yan. This allows to use the global fits to check the factorization hypothesis of the nuclear PDFs, and, indeed, these data were latter included in global fits \cite{Eskola:2009uj} providing further constrains to the badly known gluon PDF. 

From the point of view of new domains, the most interesting data appear at the forward rapidity. The typical range of $x$ measured in a  $2\to 1$ processes in $x\sim M_T\,{\rm e}^{-y}/\sqrt{s}$ \footnote{$2\to 2$ processes involve integration on a  range of $x$ leading to a larger average value. } so rather small values of $x$ can be reached at large $y$. Early predictions \cite{Albacete:2003iq}-\cite{Kharzeev:2003wz} of QCD non-linear evolution found a strong suppression of the hadronic yields with transverse momentum in the vicinity and above the saturation scale with increasing rapidity. This suppression was first observed by BRAHMS \cite{Arsene:2004ux} in a reduced range of $p_T\lesssim 4$ GeV. With improved techniques, the description of these data in terms of the CGC is rather good \cite{Albacete:2010bs}, although some issues remain to be clarified, in particular concerning the implementation of the impact parameter dependence of the nuclear wave function. However, an analysis in terms of nuclear PDFs is also possible \cite{Eskola:2008ca} but showing tension with DIS data. Additional effects as energy loss have also been put forward \cite{Frankfurt:2007rn}. In order to clarify the origin of the suppression and the relevance of saturation physics, new observables have been studied in the last years (a higher luminosity \dAu run was performed in 2008). Among them, the use of correlations could shed new light into the problem: both STAR \cite{Braidot:2010zh} and PHENIX  \cite{Adare:2011sc} have measured a suppression of the back-to-back signal in \dAu collisions when both the triggered and the associated particles are in the forward rapidities. The suppression is especially strong for central collisions. The studied $p_T$ regions are small ($p_T\lesssim2.5$ GeV for STAR and  $p_T\lesssim 1.5$ GeV for PHENIX) but interestingly the data  can be described by BK-evolved gluon distributions \cite{Albacete:2010pg}.

\section{The TeV frontiere}

The LHC nuclear program is in a situation in which no constraints for cold nuclear matter background exist from previous experiments. This includes the knowledge of the nuclear parton distributions, the nuclear effects on hadronization or even the validity of the collinear factorization. A \pPb program at the LHC will be essential to pindown these uncertainties and to  check the new domain of saturated parton densities with unprecedented precision. We present concisely arguments to support this assertion also in view of the new data presented in the QM2011 conference.

\subsection{Nuclear parton distribution functions}

\begin{figure}[h]
\begin{minipage}{0.5\textwidth}
\begin{center}
\includegraphics[width=0.8\textwidth]{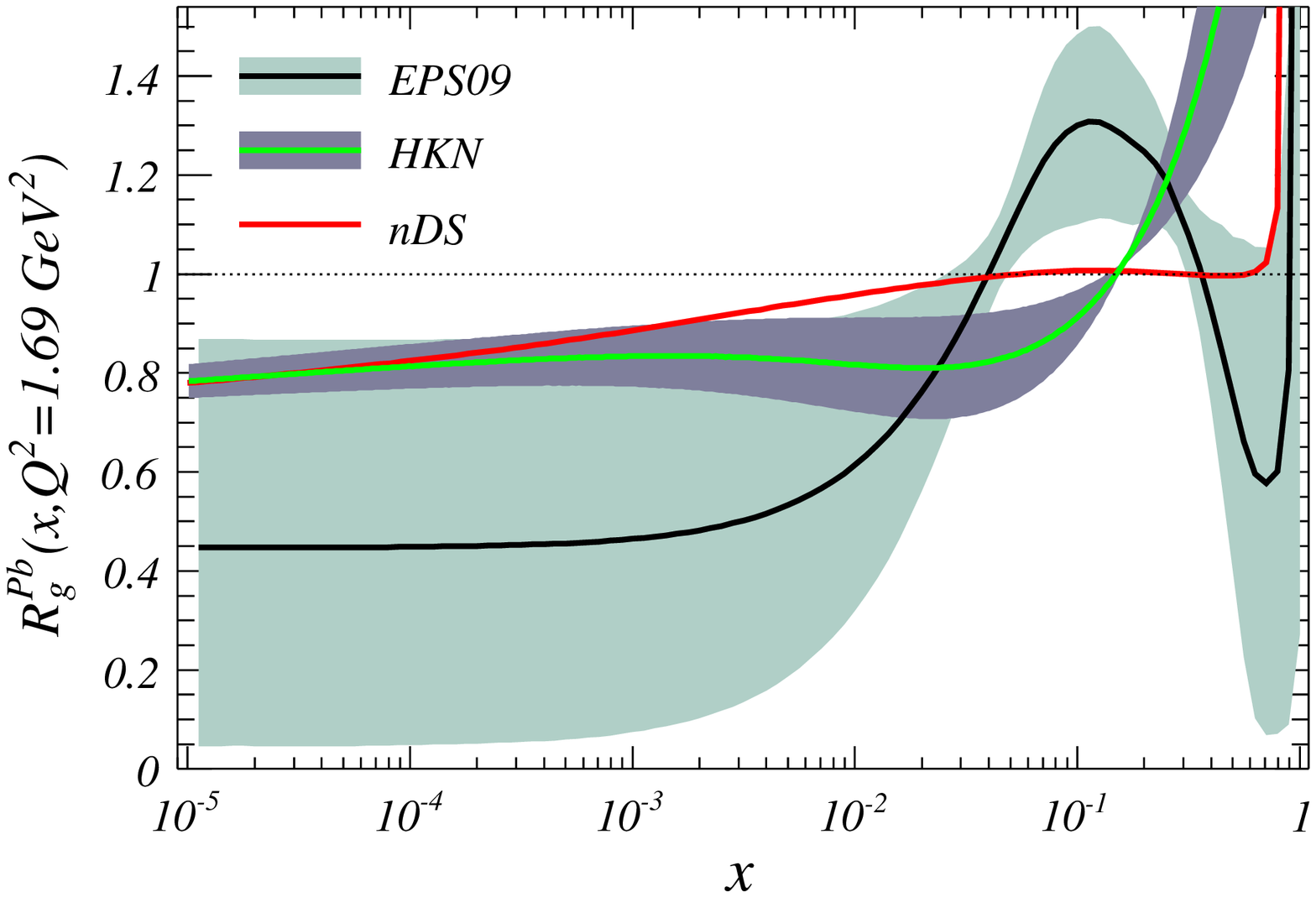}
\end{center}
\end{minipage}
\hfill
\begin{minipage}{0.5\textwidth}
\begin{center}
\includegraphics[width=0.8\textwidth]{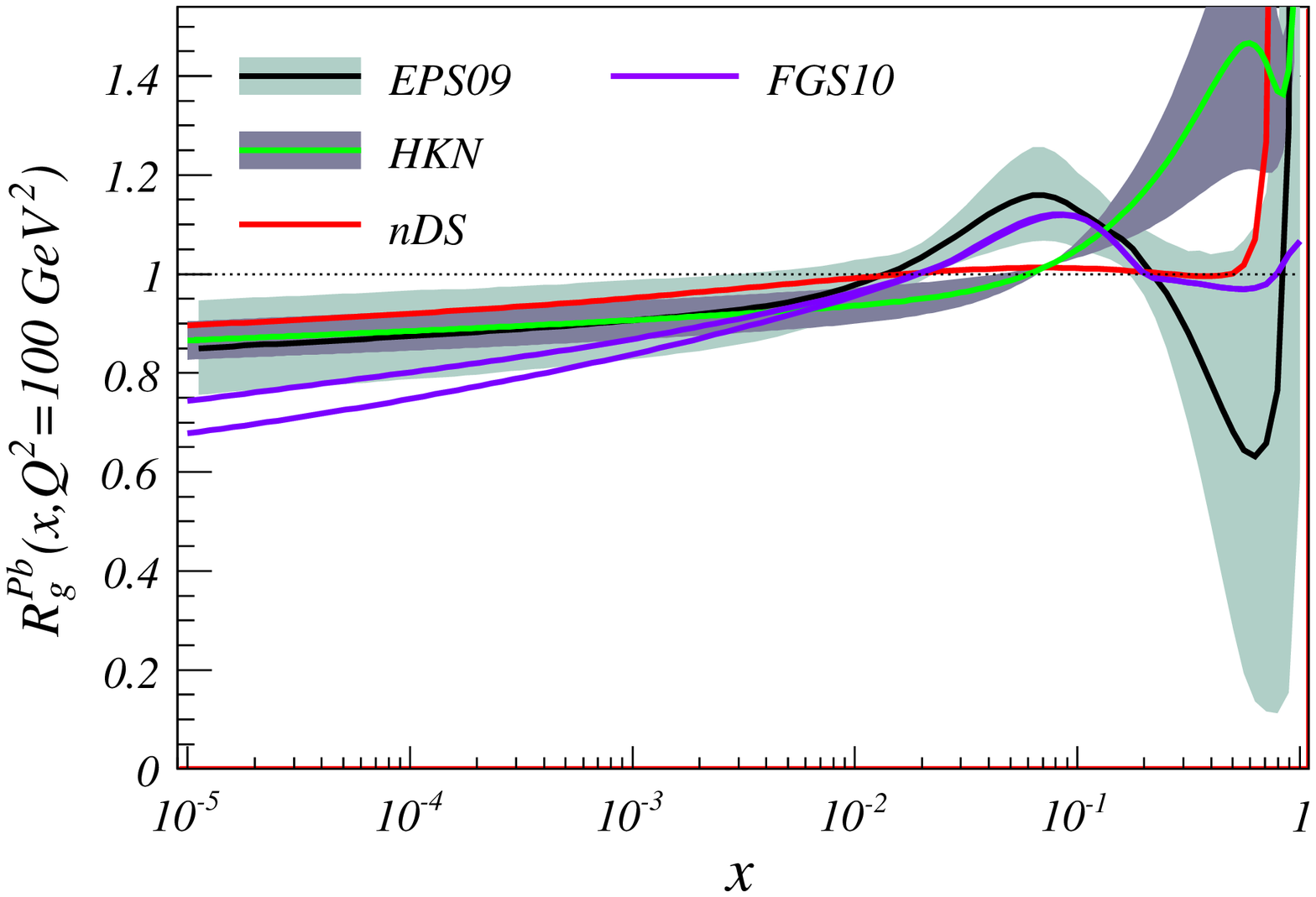}
\end{center}
\end{minipage}
\caption{Nuclear effects to the PDFs, $R_g^{Pb}(x,Q^2)$, obtained by the NLO global fits EPS09 \cite{Eskola:2009uj}, HKN07 \cite{Hirai:2007sx} and nDS \cite{deFlorian:2003qf} at two different virtualities, $Q^2=1.69$ GeV$^2$ and $Q^2$=100 GeV$^2$.}
\label{fig:npdf}
\end{figure}

The knowledge of the nuclear parton distribution functions is deficient in the needed kinematical region of the LHC, and also at forward rapidities at RHIC. This can be seen in Fig. \ref{fig:npdf} where the ratio between the gluon distributions in bound over those in free nucleons is plotted for the NLO global fits available at present. The uncertainty bands are calculated using the Hessian method and should be taken as lower limits due to remaining parametrization-bias effects.

Generically, the nuclear effects disappear at small-$x$ for both sea quarks and gluons with increasing $Q^2$, so they affect less the processes with larger virtualities, but uncertainties remain important for processes like the $J/\Psi$ production --- see below. 

The only experimental option to reduce these uncertainties in the next decade are \pA collisions. For benchmarking this is especially relevant for the LHC.

\subsection{Some process for benchmarking and nPDF studies}

\begin{figure}[h]
\begin{minipage}{0.5\textwidth}
\begin{center}
\includegraphics[width=0.8\textwidth]{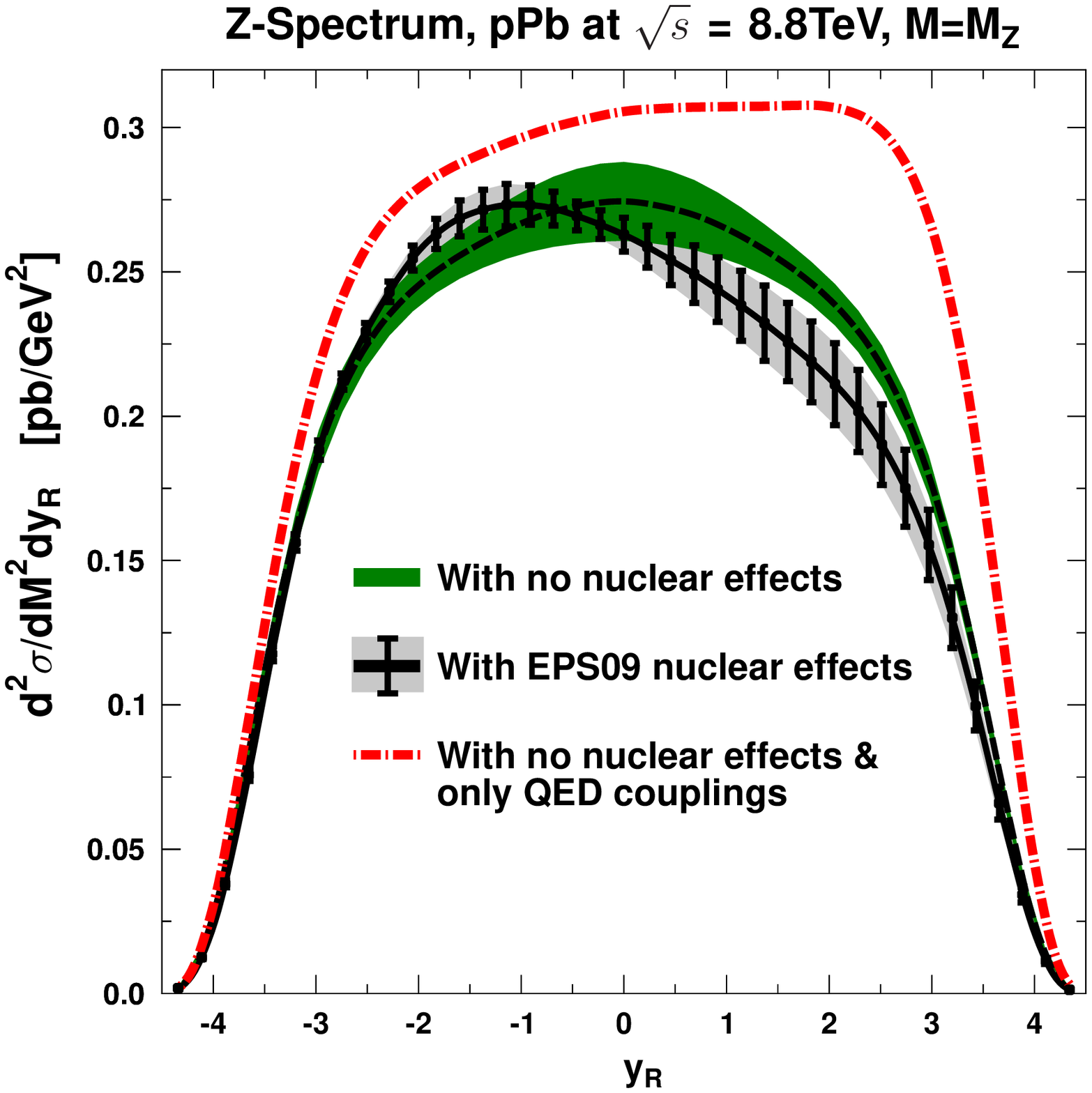}
\end{center}
\end{minipage}
\hfill
\begin{minipage}{0.5\textwidth}
\begin{center}
\includegraphics[width=\textwidth]{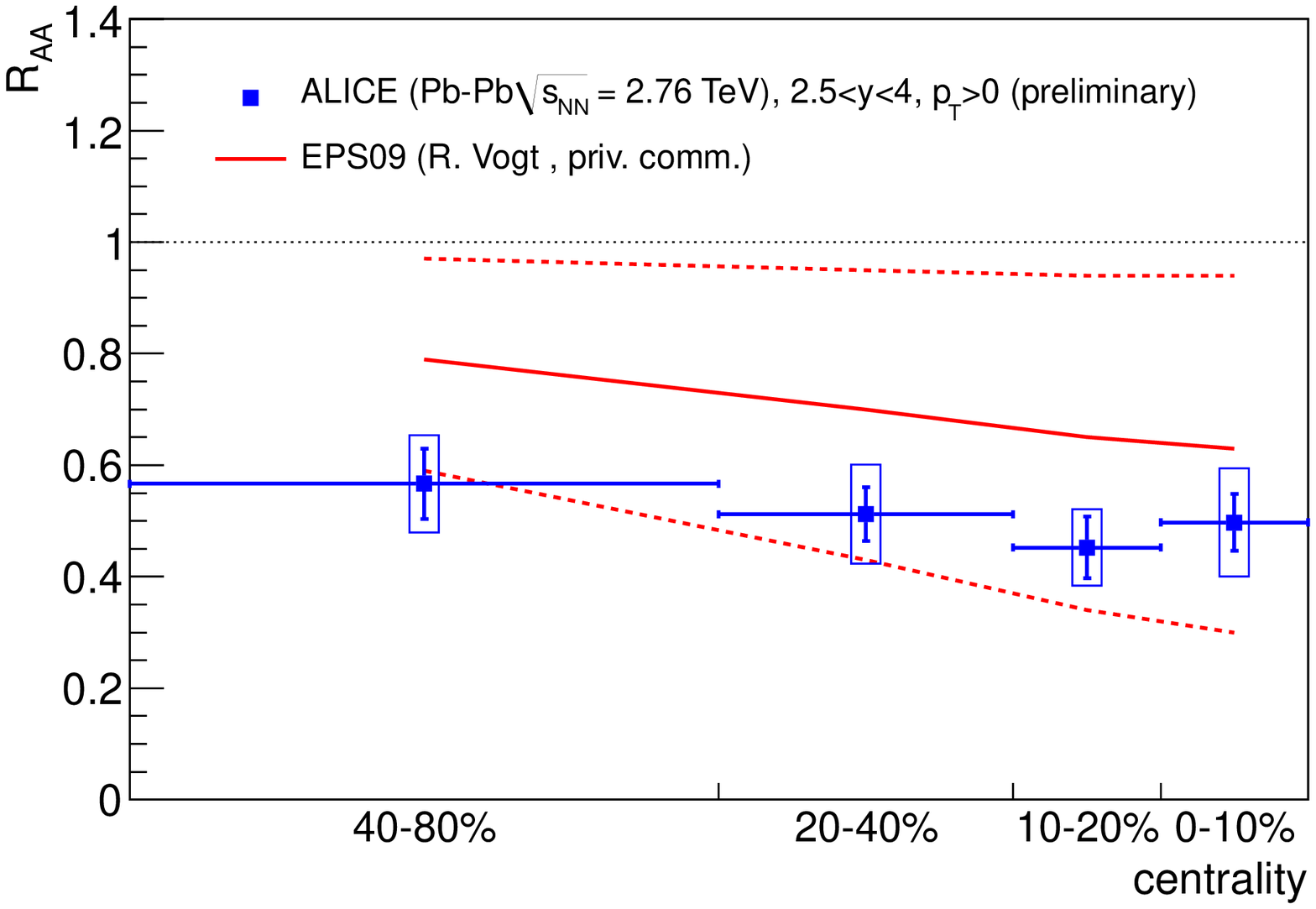}
\end{center}
\end{minipage}
\vskip -0.2cm
\caption{{\it Left}: Rapidity distributions for  dimuon pairs at the peak of the $Z$ boson in \pPb with and without nuclear effects in the PDFs \cite{Paukkunen:2010qg}. Also shown is the case for QED couplings showing larger isospin effects. {\it Right}: Suppression of the $J/\Psi$ at forward rapidities given by nuclear PDFs compared with ALICE preliminary data \cite{alice}.}
\label{fig:bench}
\end{figure}

The uncertainties in nuclear PDFs can be reduced by studying different observables both at RHIC and the LHC \cite{Salgado:2011wc}. Among them, the electroweak processes are some of the most promising ones, especially the novel available access to $W/Z$ production. One of the main advantages is that the isospin corrections are almost negligible in $Z$-production --- see Fig. \ref{fig:bench} --- making the spectrum almost rapidity-symmetric if no modification of the PDFs where present. This can be exploited to extract the nuclear effects   with small sensitivity to the proton PDFs \cite{Paukkunen:2010qg}. This is true only in asymmetric systems, as \pPb, where a forward-backward asymmetry will then directly constrain the nuclear PDFs. This feature is absent in symmetric \PbPb collisions and the corresponding sensitivity  is much more reduced.With the quoted luminosity for \pPb runs at the LHC, the statistics should be large enough for such studies. For example, 4000 $Z$'s decaying in dileptons are expected per unit rapidity in \pPb at $\sqrt{s}=8.8$ TeV, the corresponding yield for $W$ is a factor of 10 larger and the corresponding single-lepton measurements are also interesting for nPDF studies. On the other hand, process of interest as $Z$+jet would probably need a factor of 10 more luminosity for enough statistics \cite{Salgado:2011wc}. 

During the conference new data on hard processes in \PbPb at the LHC have been presented. This allows us to make first statements about the need of benchmarking and to quantify, in part, what is needed from a \pPb run. We divide the observables according to the relative amount of cold/hot QCD matter effects:

\begin{itemize}

\item {\it Processes where only cold nuclear matter effects were expected}: Interestingly, the first data on $Z$ and $W$ production are available \cite{Collaboration:2011ua,:2010px} and the agreement with pQCD factorization plus the Glauber model is excellent. More precise measurement of the luminosity and more precise data would, indeed, help to further substantiate this statement and to make the data usable for global fits. 

\item {\it Processes where both hot and cold nuclear matter could be of similar size}: Historically the paradigmatic example for this second kind of processes is the quarkonia suppression in nuclear collisions. From SPS to RHIC and now to the LHC, a strong suppression of the $J/\Psi$ was measured --- now also of different quarkonia states, including $\Upsilon$'s \cite{Chatrchyan:2011pe}. The corresponding cold nuclear matter effects measured in \pA and \dAu collisions extrapolated to \PbPb or \AuAu lead to suppression of magnitude of the same order as the one observed, so the interpretation of the data requires quantifying these effects with enough precision. For the LHC, the first data have been presented \cite{alice,quarkonia} in the conference. They are taken in different kinematical domains and for different particle species. A complete picture of the suppression will indeed need of a good control over the cold nuclear matter effects. Just to mention two examples: (i) Taken at face value, the suppression found by ALICE \cite{alice} for the $J/\Psi$ in the forward region is in agreement with the suppression due to shadowing as given by the EPS09 parametrization with no need of extra hot matter effects --- see Fig. \ref{fig:bench}; (ii) CMS measure the excited $\Upsilon$ states  to be more suppressed than the  $\Upsilon$  ground state \cite{Chatrchyan:2011pe}. Old data from \pA  \cite{Alde:1991sw} showed, however, no different suppression. This would point to a genuine hot matter effect measured by CMS. Clearly further checks  with \pPb collisions at the LHC energies and with high statistics are of utmost importance for the  physical picture of quarkonia suppression.

\item {\it Processes where cold  matter effects are expected to be much smaller than hot matter effects.} The main example is in this case the production of particles with high transverse momentum, including the reconstructed jets. This has been one of the main observables discussed at the Quark Matter conference \cite{jetquenching}. At RHIC, as commented before, the nuclear PDFs  were already explored by previous experiments for the data at central rapidities. At the LHC this is, however, not the case and, although the bulk of the effect is expected to be due to {\it jet quenching} a precise interpretation would need a better knowledge of the cold nuclear matter effects, which could, for example, change the slope of the single-particle suppression $R_{AA}$. On the other hand, new data from \dAu at RHIC \cite{dAujets} seems to indicate a modification of the jet structures. Some of the theoretical developments in the last year \cite{MehtarTani:2011tz,MehtarTani:2010ma,MehtarTani:2011jw,CasalderreySolana:2011rz}  point to the decoherent gluon emission off the different emitters in the shower as a  new relevant contribution. This decoherence is a very generic phenomena which could affect also jets in  \pA. More theoretical work is indeed needed but a  measurement of reconstructed jets in \pPb  would be welcome. 

\end{itemize}

\subsection{The saturation of partonic densities}

The novel kinematical regimes open for QCD studies in nuclear collisions allow for an unprecedented access to the small-$x$ region of the wave function \cite{javier}. Rather general arguments imply that the linear regime in the evolution equations of the parton distributions should cease to be valid at a scale $Q_{\rm sat}^2\simeq A^{1/3}x^{-\lambda}$ in which the proportionality factor and the exponents are not completely known. In the last  decade  a whole formalism to systematically include these corrections in different approximations has been developed and given the generic name of Color Glass Condensate (CGC) \cite{javier}. The main realization is a set of evolution equations which are known at NLO and which generalize the Balitsky-Fadin-Kuraev-Lipatov equations to the non-linear case. These Balitsky-Kovchegov (BK) equations at NLO provide the first theoretically controlled tool to apply the formalism in experimental situations and to perform precision checks of the relevance of non-linear terms in the evolution equations \cite{paloma}.

The presence of a scale, $Q_{\rm sat}^2$, which for large enough nuclei or small enough $x$ could be in the perturbative region, allows to perform calculations of quantities usually considered to belong to the  {\it soft} physics using small coupling techniques. For example, the total multiplicities in nuclear collisions can be described by the solution of the running coupling BK equations. Nucleus-nucleus collisions is, however, not the best experimental condition to study the physics of saturation due to the presence of final state hot QCD matter effects which effectively wash-out most of the information about the nuclear wave function, except perhaps global quantities as multiplicities or some long range correlations. In this situation, \pA collisions  offer the best experimental conditions in the next decade before a possible lepton-ion collider as the EIC or the LHeC are built \cite{stasto}. Moreover, the new techniques to study the initial stages of the collision by methods of the CGC call for a benchmarking of the initial wave function of the nucleus for a precise determination of quantities as e.g. the viscosity of the produced medium. In this sense the benchmarking role of the \pA which was traditionally limited to the hard probes is nowadays of much wider scope \cite{javier}.

\section{Summary and perspectives}

Proton-nucleus collisions offer a variety of physics to be studied as well as the needed benchmarking conditions for cold nuclear matter background subtraction in nucleus-nucleus. Preliminary studies confirm the feasibility of \pA collisions in the special configuration of the LHC \cite{Salgado:2011wc}. The jump in energy will be a factor of more than 40 at top LHC energy or more than 20 at present energies that of RHIC \dAu collisions. The phase space region open for new studies is probably the largest one in the history of particle physics for the same colliding system. Interestingly, this is also one of the rare times in which a large overlap in kinematic regimes is available in two parallel nuclear programs. This is an ideal situation to reduce the uncertainties by comparing the data in two different experimental situations and the benefits for nPDF  and/or saturation studies cannot be overemphasized. 

For the particular case of the LHC, the goals presented in this paper cannot be reached with \PbPb collisions alone.

\section*{Acknowledgments}

I would like to thank the authors of the report on proton-nucleus collisions at the LHC, Ref.  [1], at the origin of most of the material presented here. I would also like to thank Enrico Scomparing for providing me with Figure 3 Right. This work is supported by Ministerio de Ciencia e Innovacion of Spain, Xunta de Galicia, by project Consolider-Ingenio 2010 CPAN and Feder. CAS is a Ram\'on y Cajal researcher.

\end{document}